# Electron as a Complex-Dynamical Interaction Process

Andrei P. Kirilyuk[*]

Institute of Metal Physics, Kiev-142, Ukraine 03142

## Abstract

A system of two initially homogeneous, physically real fields uniformly attracted to each other is considered as the simplest basis of the self-developing world structure. It is shown that, for generic interaction parameters, the system is unstable against periodic cycles of self-amplified, essentially nonlinear squeeze of its extended part to a small volume around randomly chosen centre, followed by the reverse extension. The resulting spatially random pulsation, or "quantum beat", is observed as (massive) elementary particle such as the electron. The property of mass is then universally and consistently defined as temporal rate of such dynamically chaotic and essentially nonlinear quantum beat, without introduction of any additional entities. The chaotic dynamical structure of the electron is based on the qualitatively new phenomena of dynamic redundance (multivaluedness) and entanglement of the interacting protofields, revealed due to the unreduced, universally nonperturbative interaction analysis. The obtained picture can be considered as complex-dynamical completion of the "double solution" concept proposed by Louis de Broglie. The dynamically emerging wave-particle duality, quantum discreteness, indeterminacy, space, and time lead to the equations of special relativity and quantum mechanics, providing their causal explanation and explicit unification. The elementary particle structure, its intrinsic properties, quantum and relativistic behaviour are obtained thus all together, within the unified analysis of the unreduced interaction process leading to the universal concept of dynamic complexity. The same complex-dynamical process accounts for the universal gravitation and general relativity, unifying it dynamically with the extended quantum mechanics and special relativity. The electromagnetic, weak, and strong types of interaction between particles also constitute integral, dynamically unified parts of quantum beat processes inside elementary particles. The classical, dynamically localised behaviour emerges in a closed system as a higher complexity level corresponding to formation of elementary bound systems (like atoms).

---

[*] Address for correspondence: Post Box 115, Kiev-30, Ukraine 01030. E-mail address: kiril@metfiz.freenet.kiev.ua.

CONTENTS

# 1. Introduction

Physical and mathematical incompleteness of existing interpretations of quantum behaviour is a well-known fact explicitly emphasised as a basic feature within the standard, or Copenhagen, formulation of quantum mechanics [1-5]. The fundamentally inexplicable, "mysterious" character of the essentially quantum properties, such as wave-particle duality and quantum indeterminacy, is disputed within a group of so-called causal approaches, trying to reduce the number of specific "quantum" postulates to a general level of classical physics [4]. Note, however, that those approaches do not include any detailed study on the *physical structure* of elementary particles as such, simply postulating its either point-like or (linear) wave-like model (or a "mysterious" combination of both). Such separation between "structure" and its "function" (i. e. properly "quantum" behaviour of the particle) cannot be correct, since both belong to the same level of dynamics of the same physical system, but it remains a forced simplification within the traditional, "linear" science approach which has the well-known fundamental difficulties in description of the unreduced, essentially nonlinear structure-formation processes.

The first serious attempt to include the dynamically changing particle structure and its causally understood quantum behaviour into the same picture was realised by Louis de Broglie within his intrinsically nonlinear "wave mechanics" [6-10]. It is proposed in the form of the "double solution" [7-9] further amplified by the "thermodynamics of the isolated particle" [10] and trying to unify formal representations of the undulatory and corpuscular aspects of quantum behaviour into a unique, explicitly presented and reality-based solution, which inevitably includes the essential clarification of the physical particle structure.[*)] However, despite a number of fundamentally important results obtained within the unifying logic of the original double solution concept, including the rarely reproduced detailed derivation of the standard expression for the de Broglie wavelength [6], de Broglie's project in the whole has not been completed.

In this paper we present an independent, first-principles analysis of complex (chaotic) dynamics of a system of two *a priori* homogeneous physical fields, or *protofields*, attracted to each other and chosen as the simplest possible basis of the emerging world structure. We show that the explicitly obtained solution of a new, intrinsically complete type naturally unifies, simply due to its unreduced (nonperturbative) character, all the expected properties of de Broglie's double solution and thus provides the consistent solution of "quantum mysteries". This is related to the qualitatively new phenomenon of *dynamically redundant (multivalued) entanglement* of interacting entities that naturally emerges within the proposed universally

---

[*)] In this respect, the double solution should be clearly distinguished from its reduced, schematic version first proposed by Louis de Broglie himself under the name of "pilot-wave interpretation", reintroduced much later by Bohm [11], further developed by his followers [4,12,13], and now often presented as the unique version of "causal de Broglie-Bohm approach" (see also refs. [14,15]).

nonperturbative analysis (section 2) and constitutes the basis of the recently advanced, reality-based concept of dynamic complexity in any nontrivial system with interaction [14-17]. The explicitly creative, structure-forming character of the obtained solution automatically provides the internal physical structure of the electron and other emerging elementary particles, each massive particle being explicitly obtained thus as a chaotic (dynamically multivalued) interaction process that takes the form of spatially chaotic, highly nonlinear pulsation, or *quantum beat* [14-16], developing in an a priori totally deterministic (and ultimately simple) system of coupled protofields (section 3).

Due to the absolute universality of the analysis and the obtained general solution, the specific "quantum" mysteries are causally resolved within the same approach that provides complex-dynamical completion of the canonical description of higher-level, classical phenomena emerging in a natural hierarchy of further interaction development between the explicitly obtained entities of the preceding (lower) complexity levels, which is important for unification of science around the causally complete understanding of unreduced diversity of the complex world dynamics [14]. Thus, the unreduced, dynamically multivalued dynamics of the interacting elementary particles gives rise to the genuine quantum chaos and causal quantum measurement phenomena, as well as to the classical (permanently localised) type of behaviour (section 7) dynamically emerging in a *closed* quantum system as the next (higher) level of unreduced dynamic complexity, in the form of the simplest bound system (like atoms) [14-17]. The number and properties of the "four fundamental forces" of interaction between particles also naturally emerge in the described world configuration of two coupled protofields (section 6), where all the four forces are intrinsically, dynamically unified within quantum beat processes (i. e. "within" elementary particles).

Moreover, the quantum beat pulsation provides not only the causally specified internal structure of elementary particles, but also physically real and "emerging" (rigorously derived) space and time, in direct relation to the global particle motion, which leads to the causally derived, totally dynamic, and unified understanding of the conventional "relativistic" relations between space, time, motion, mass, energy, other properties, and their genuine physical meaning in terms of the underlying complex (multivalued) dynamics (section 4). One obtains, in that way, the intrinsic, dynamic unification between causally extended versions of quantum mechanics and relativity (sections 4, 5), persistently missing in the conventional, dynamically single-valued (unitary) theory. This unification includes also the irreducibly quantum origin of gravity and causally extended "general relativity", representing but particular aspects of the holistic complex dynamics of interacting particle-processes within the single system of two interacting protofields (section 6).

It becomes clear that such unified, totally causal (realistic) solution of the conventional quantum (and other) "mysteries" is based essentially on the proposed unreduced, universally nonperturbative analysis of any interaction process revealing the phenomenon of dynamic multivaluedness and entanglement, whereas the conventional theory always simplifies the underlying system interaction by using one or another version of perturbation theory, which leads inevitably to a fatally reduced, dynamically single-valued "model" of reality that cannot reproduce its observed properties without dramatic contradictions, emerging in the form of "inexplicable", postulated mysteries (and entities) and persistent, "irresolvable" problems. Below we briefly specify the difference between these two pictures (reduced and unreduced) of world dynamics, starting from the dynamical structure of electron and other elementary particles that emerge *together* with their quantum and relativistic behaviour, mass-energy and other intrinsic properties.

## 2. Complex dynamics of interaction between two protofields: dynamic reduction, entanglement, and causal randomness

Consider a system of two physically real and initially homogeneous protofields, or media, one of them being associated by its physical quality with eventually emerging electromagnetic (e/m) phenomena and the other with gravitational effects, and let us analyse the results of their attractive interaction, or coupling, trying to avoid any perturbative reduction and additional assumption. It is implied that the protofields may constitute the fundamental — and the simplest possible — physical basis of the *explicitly emerging* world structure, if we can show how all the observed features of the latter, starting from the elementary particles, progressively and naturally form in course of autonomous, or "spontaneous", development of the unreduced interaction process. Indeed, a system of only two a priori homogeneous and homogeneously interacting entities constitutes the simplest possible physical basis of a viable world, since a still simpler configuration without interaction cannot provide any system development (structure emergence), while attribution of a particular physical nature to the e/m and gravitational protofields forms the indispensable minimum of (justified) "postulates" of the theory. It is also evident that the free protofields, in their structureless initial state, actually provide the causal, non-contradictory prototype of "ether", since they cannot be observed as such from the inside of the emerging world, but remain physically real (which is confirmed by observability of their perturbations, resulting from interaction).

The state of the compound system of interacting fields should be represented by the state-function, $\Psi(q,\xi)$, satisfying a dynamical equation, called here existence equation, which effectively generalises various model equations and can eventually be specified, but initially serves simply to fix the fact of unreduced, inseparable protofield interaction within the integrated (dynamically mixed) system:

$$\left[ h_\text{e}(q) + V_\text{eg}(q,\xi) + h_\text{g}(\xi) \right] \Psi(q,\xi) = E \Psi(q,\xi), \tag{1}$$

where $q$ denotes the electromagnetic and $\xi$ the gravitational degrees of freedom, $h_\text{e}(q)$ and $h_\text{g}(\xi)$ are the generalised Hamiltonians for the free (non-interacting) media (i. e. any actually measured functions representing a measure of dynamic complexity defined below), $V_\text{eg}(q,\xi)$ is an arbitrary (though eventually attractive and binding) interaction potential between the fields of $q$ and $\xi$, and $E$ is the energy (or other compound-system property corresponding to the selected generalised Hamiltonian).

To analyse the interaction development, we expand $\Psi(q,\xi)$ over the complete system of eigenfunctions, $\{\phi_n(q)\}$, of the free e/m protofield Hamiltonian, $h_\text{e}(q)$:

$$h_\text{e}(q)\phi_n(q) = \varepsilon_n \phi_n(q), \tag{2}$$

$$\Psi(q,\xi) = \sum_n \psi_n(\xi)\phi_n(q), \tag{3}$$

which transforms eq. (1) into the equivalent system of equations:

$$\left[ h_\text{g}(\xi) + V_{00}(\xi) \right]\psi_0(\xi) + \sum_n V_{0n}(\xi)\psi_n(\xi) = \eta \psi_0(\xi), \tag{4a}$$

$$\left[ h_\text{g}(\xi) + V_{nn}(\xi) \right]\psi_n(\xi) + \sum_{n'\neq n} V_{nn'}(\xi)\psi_{n'}(\xi) = \eta_n \psi_n(\xi) - V_{n0}(\xi)\psi_0(\xi), \tag{4b}$$

where
$$\eta_n \equiv E - \varepsilon_n \,, \tag{5}$$

$$V_{nn'}(\xi) = \int_{\Omega_q} dq \phi_n^*(q) V_{eg}(q,\xi) \phi_{n'}(q) \,, \tag{6}$$

and we have separated the equation with $n = 0$, while assuming that $n \neq 0$ in other equations and $\eta \equiv \eta_0$. Note that the obtained system of equations, eqs. (4), is just another, more relevant form of the same existence equation, eq. (1), which does not involve any additional assumption or result of interaction development. Such expression of the system configuration in terms of dynamical "eigen-modes" (or "elements") of free components should always be possible, even for the formally "nonlinear" protofields, since we suppose that the internal dynamics of the latter is known, or "integrable", at least in the range of participating scales (or else one should start with a similar analysis of unknown lower-level dynamics).

Expressing $\psi_n(\xi)$ from eqs. (4b) with the help of the standard Green function technique [18,19] and inserting the result into eq. (4a), we reformulate the problem in terms of effective existence equation formally involving only gravitational degrees of freedom ($\xi$):

$$\left[ h_g(\xi) + V_{eff}(\xi;\eta) \right] \psi_0(\xi) = \eta \psi_0(\xi) \,, \tag{7}$$

where the effective (interaction) potential (EP), $V_{eff}(\xi;\eta)$, is given by

$$V_{eff}(\xi;\eta) = V_{00}(\xi) + \hat{V}(\xi;\eta) \,, \quad \hat{V}(\xi;\eta) \psi_0(\xi) = \int_{\Omega_\xi} d\xi' V(\xi,\xi';\eta) \psi_0(\xi') \,, \tag{8a}$$

$$V(\xi,\xi';\eta) = \sum_{n,i} \frac{V_{0n}(\xi) \psi_{ni}^0(\xi) V_{n0}(\xi') \psi_{ni}^{0*}(\xi')}{\eta - \eta_{ni}^0 - \varepsilon_{n0}} \,, \quad \varepsilon_{n0} \equiv \varepsilon_n - \varepsilon_0 \,, \tag{8b}$$

and $\{\psi_{ni}^0(\xi)\}$, $\{\eta_{ni}^0\}$ are the complete sets of eigenfunctions and eigenvalues for an auxiliary, truncated system of equations (where $n, n' \neq 0$):

$$\left[ h_g(\xi) + V_{nn}(\xi) \right] \psi_n(\xi) + \sum_{n' \neq n} V_{nn'}(\xi) \psi_{n'}(\xi) = \eta_n \psi_n(\xi) \,. \tag{9}$$

The general solution of the initial existence equation, eq. (1), is then obtained as:

$$\Psi(q,\xi) = \sum_i c_i \left[ \phi_0(q) + \sum_n \phi_n(q) \hat{g}_{ni}(\xi) \right] \psi_{0i}(\xi) \,, \tag{10}$$

$$\psi_{ni}(\xi) = \hat{g}_{ni}(\xi) \psi_{0i}(\xi) \equiv \int_{\Omega_\xi} d\xi' g_{ni}(\xi,\xi') \psi_{0i}(\xi') \,, \quad g_{ni}(\xi,\xi') = V_{n0}(\xi') \sum_{i'} \frac{\psi_{ni'}^0(\xi) \psi_{ni'}^{0*}(\xi')}{\eta_i - \eta_{ni'}^0 - \varepsilon_{n0}} \,, \tag{11}$$

where $\{\psi_{0i}(\xi)\}$ are the eigenfunctions and $\{\eta_i\}$ the eigenvalues found from the effective equation, eq. (7),

while the coefficients $c_i$ should be determined from the state-function matching conditions along the boundary where interaction vanishes. The observed system density, $\rho(q,\xi)$, is given by the squared modulus of the protofield amplitude: $\rho(q,\xi) = |\Psi(q,\xi)|^2$.

The described equivalent problem transformation is known as the standard "optical potential method" [19] in which, however, it is invariably followed by a perturbative reduction of the EP expression, eqs. (8), eliminating the "nonintegrable" nonlinear dependence on the unknown eigen-solutions, but simultaneously "killing" all the intrinsic dynamic complexity it hides. Indeed, as shown below (see also refs. [14-18]), in its unreduced form this self-consistent EP dependence on the eigen-solutions to be found represents the *essential, dynamic nonlinearity* of a generic, often a priori "linear" interaction process, in the form of *dynamically emerging* feedback loops of interaction, which leads to a dramatic increase of the number of solutions, each of them being "complete" in the ordinary sense and thus incompatible with any other, equally real, solution. This phenomenon of *dynamic redundancy, or multivaluedness*, of the unreduced problem solution can be demonstrated directly, by counting the number of eigen-solutions of eq. (7) determined by the highest power of the eigenvalue $\eta$ in the corresponding characteristic equation. If $N_\text{e}$ and $N_\text{g}$ are the numbers of eigen-modes (or "elements") for the free e/m and gravitational protofields respectively (normally $N_\text{e} = N_\text{g}$ and $N_\text{e}, N_\text{g} \gg 1$), then the total number of eigenvalues for eq. (7) is

$$N_\text{eg} = N_\text{g} N^1_\text{eg} = N_\text{g}(N_\text{e} N_\text{g} + 1), \qquad (12)$$

where the factor of $N^1_\text{eg} = N_\text{e} N_\text{g} + 1$ is due to the EP dependence on $\eta$, eq. (8b), providing the $N_\text{g}$-fold redundancy of the ordinary complete sets of $N_\text{e} N_\text{g}$ eigen-solutions and a separate, reduced set of $N_\text{g}$ eigen-solutions that forms a specific, "intermediate" system state specified below.

Whereas other, equivalent ways of estimation of the number of solutions of eq. (7) only confirm this result and show that all the solutions obtained for generic interaction parameters are real (not spurious) [14-18], there is a still more transparent, physical support for common emergence of dynamic redundancy. It comes from the fact that within the unreduced interaction process every mode of each partner interacts, by physical, dynamic entanglement (or mixing), with every mode of other partner(s), which should give $N_\text{e} N_\text{g} = (N_\text{g})^2$ entangled configurations in our case. However, the number of "places" in reality for them remains the same as for each of the free interaction partners (i. e. equal to $N_\text{e} = N_\text{g}$), since the products of interaction exist in the same fundamental reality as the initial objects, and this gives the $N_\text{g}$-fold redundancy. The conventional, actually always perturbative interaction analysis corresponds to a "weak" contact of each element of an interacting entity with only one element of another interaction partner, so that all those redundant, real solutions are lost, except the single, trivial one, which explains the dramatic reduction of reality within the conventional theory, crudely simplifying the interaction result and leading, in particular, to the inexplicable "mysteries" of the standard, unitary quantum mechanics, within any its speculative "interpretation" (we continue to specify this conclusion below).

We call each of the complete, and therefore mutually *incompatible*, solutions of the unreduced effective problem formulation, eqs. (7)-(11), *realisation* of the system and shall enumerate realisations by index $r$ ($r = 1, 2, ..., N_\text{g}$). For the considered case of interacting, initially homogeneous protofields each regular realisation is represented by a dynamically auto-squeezed, or (physically) "reduced" state of the fields concentrated around thus emerging centre of reduction, or physical "space point" (it represents also the localised, "corpuscular" state of the thus emerging elementary field-particle, see below). This can be seen directly from eqs. (7)-(11), if we choose the starting expansion basis in eq. (3), $\{\phi_n(q)\}$, in the form of a complete system of highly localised functions, close to $\delta$-functions. Then the eigenvalues, like $\varepsilon_n$ and $\eta_i$, will correspond to "point coordinates" for the eventually emerging space, and it can be seen from eqs. (11) that the state-function $\Psi_r(q,\xi)$ for each $r$-th realisation is concentrated around certain "preferred" point/eigenvalue, $\eta_i^r$ due to the resonant structure of denominators (in its relation to the "cutting" numerators with "overlap integrals"). In order to better see this nonperturbative, *essentially nonlinear* effect of "dynamical auto-squeeze" in the homogeneously interacting protofield system, let us write down the explicit expression for the density and state-function of the system in its $r$-th realisation obtained by substitution of the $r$-th eigen-solution set, $\{\psi_{0i}^r(\xi), \eta_i^r\}$ (found from the unreduced effective existence equation, eq. (7)), into eqs. (10)-(11):

$$\rho_r(q,\xi) = |\Psi_r(q,\xi)|^2,$$

$$\Psi_r(q,\xi) = \sum_i c_i^r \left[ \phi_0(q)\psi_{0i}^r(\xi) + \sum_{n,i'} \frac{\phi_n(q)\psi_{ni'}^0(\xi) \int_{\Omega_\xi} d\xi' \psi_{ni'}^{0*}(\xi')V_{n0}(\xi')\psi_{0i}^r(\xi')}{\eta_i^r - \eta_{ni'}^0 - \varepsilon_{n0}} \right]. \quad (13)$$

Due to the nonperturbative, effectively nonlinear dependence of the unreduced solution on the problem eigen-solutions (in the terms of the inner sum in eq. (13)), the system density in the $r$-th realisation will indeed be concentrated around the corresponding "centre of reduction", $\eta = \eta_i^r$, for certain $i$ (it is convenient to enumerate the eigenvalues so that $i = r$ for this particular $i$). This dynamic state-function/density concentration is closely related to the dynamic formation of the corresponding EP potential well in the previously homogeneous system configuration just around the same preferred physical "point", $\eta = \eta_r^r$, so that both these aspects of the dynamic reduction (collapse) of the system form the single, self-sustained process, in agreement with the effective form of the existence equation, eq. (7). The second, EP aspect of the nonlinear reduction directly follows from the nonperturbative EP expression, eq. (8), which can be presented for the considered $r$-th system realisation as

$$V_{\text{eff}}(\xi;\eta_i^r)\psi_{0i}^r(\xi) = V_{00}(\xi)\psi_{0i}^r(\xi) + \sum_{n,i'} \frac{V_{0n}(\xi)\psi_{ni'}^0(\xi) \int_{\Omega_\xi} d\xi' \psi_{ni'}^{0*}(\xi')V_{n0}(\xi')\psi_{0i}^r(\xi')}{\eta_i^r - \eta_{ni'}^0 - \varepsilon_{n0}}, \quad (14)$$

It is important here that the unreduced EP has the same resonant, nonlinear dependence on the eigen-solutions as the state-function of eq. (13), and therefore will dynamically form the potential well just around the same centre of localisation (which is described mathematically by the same resonant and cutting structures in eq. (14)) [14-17].

The dynamic auto-squeeze (reduction) process of a system towards one of its realisations from the initial structureless state provides the universal mechanism of *any* structure formation [14], considerably extending the conventional "self-organisation" concept (based on the same single-valued, perturbative description as the canonical "linear" science and thus actually needing artificial insertion of the "self-organised" structure into the main equations supposed to "naturally produce" it). The dynamic reduction process shows how the system actually, physically "produces" and "takes" one of its incompatible realisations by following a dynamically self-amplifying, catastrophic (avalanche-type) phase of the (generic) interaction process that possesses thus a permanent, omnipresent instability driven by the unceasingly forming feedback loops of the unreduced interaction. This universal dynamic instability of an "ordinary" interaction has a very transparent physical representation completing the equally simple interpretation of dynamic redundance described above and confirming the rigorous analysis of eqs. (13)-(14). Indeed, it is quite clear that the more is the concentration of each protofield around certain locality (starting from some initial, in principle infinitesimal, fluctuation), the more is their mutual attraction at that locality yet increasing field concentration, etc. Correspondingly, the conventional, perturbative analysis always applied within the canonical, unitary theory can be described as incorrect cutting of those emerging interaction loops, which leads to artificial interruption of dynamic instability and the related structure formation, accompanied by (unjustified) reduction of the dynamic multivaluedness of that process to only one, slightly inhomogeneous solution-realisation. Therefore it is only the unreduced, dynamically multivalued and interaction-driven entanglement between the interacting protofields, described by the nonperturbative EP formalism, eqs. (7)-(14), that provides the adequate picture of the internal dynamical structure of the electron and other (massive) elementary particles, where the above dynamical squeeze (reduction) of the protofields accounts for the "corpuscular", localised state formation within the total field-particle dynamics (further details are considered below).

It is clear also that the described dynamic, self-sustained structure formation through the autonomous interaction feedback development constitutes the property of real, and universally defined, *nonlinearity* of *any* generic system with interaction that is not explicitly inserted into the initial problem formulation, as it is always done in the conventional, single-valued imitations of a loosely understood "nonlinearity" (which is introduced "heuristically", in the form of "curved", or "nonlinear", functional dependencies, including artificially inserted "recursive", self-acting equation terms, "rugged landscapes" in abstract "spaces", numerically obtained "attractors", etc.). We call the unreduced, *dynamically* emerging (interaction-driven) and universal "self-action" property of dynamical systems, eqs. (7)-(11), the *essential, or dynamic, nonlinearity* [14] and state that any other meaning of nonlinearity inevitably suffers from incompleteness and ambiguity. The essential nonlinearity provides, in the case of protofield interaction, the causally complete version of that "hidden", intrinsic nonlinearity of an externally "linear" interaction process (cf. eq. (1)) which is indispensable for the consistent understanding of quantum mechanics, as it was repeatedly emphasised by Louis de Broglie [9] (we further specify this statement below). We have shown above that the same unreduced interaction process that determines the essential nonlinearity of any real system is at the origin of dynamic multivaluedness of its emerging realisations. Therefore any dynamically single-valued, effectively one-dimensional (or "separable") description of the conventional analysis can be characterised as *essentially linear* one, irrespective of the externally "curved" shape of an "exact" solution (such as the canonical soliton) or formally postulated system "stochasticity", which is always equivalent to a dynamically single-valued projection of the real system dynamics.

Since the system has *many* equally possible realisations (due to the essential nonlinearity of the unreduced interaction process), and the driving interaction never stops its action, any system realisation cannot be stable and the system will start leaving it, within the same interaction process, immediately after the end of the corresponding reduction phase. Indeed, each of the protofields in their squeezed state within a given (currently occupied) realisation continues to interact with the surrounding, "non-squeezed" portions of its interaction partner (the other protofield). In the stage of catastrophic reduction these "sideways" interactions are dynamically suppressed by the main self-amplifying tendency, but after the latter has produced the maximum possible protofield compression (the physically real protofields have a finite compressibility) and stopped, the "main", dynamically self-amplifying tendency is reversed and the system catastrophically extends itself, "drawn out" by the surrounding "flat" portions of the protofields. After having attained the completely extended, quasi-free state, the local configuration of the interacting protofields will evidently repeat the squeeze/reduction phase, but towards a new "point"-realisation (centre of reduction) "chosen" by the system in the *causally random* fashion among the adjacent multiple, equally possible realisations. In this way, the essential nonlinearity (dynamic instability) in the a priori homogeneous system of interacting protofields leads to "spontaneous" (purely dynamic) emergence of a highly nonlinear "clock-work", producing unceasing series of cycles of alternating local protofield squeeze (reduction) and extension. This fundamental, naturally "quantized" process of essentially nonlinear and spatially chaotic pulsation, emerging eventually around as many locations as there are "elementary particles" in the coupled protofield universe, is called *quantum beat* [14-16] and provides the world with the physically real, emerging entities of *time* and *space* that were not present at all in the starting existence equation, eq. (1). Quantum beat constitutes also the essence of the internal dynamical structure of the electron and other massive elementary particles, and we continue to specify it below.

The reality of that complex dynamical structure of the externally simple configuration of two interacting protofields is supported by the dynamically squeezed, highly inhomogeneous configuration of the unreduced, nonperturbative solution for each "regular" system realisation considered above, eqs. (13)-(14), but also by the opposite, quasi-homogeneous (quasi-free) structure of the exceptional, transitional state/realisation between the regular realisations that corresponds to the extension phase of the quantum beat process and is explicitly given by the same nonperturbative version of the EP formalism. Indeed, it is easy to see that the effective existence equation, eqs. (7)-(8), always accepts one and only one particular solution-realisation with quasi-homogeneously distributed density that corresponds to a really weak addition, $\hat{V}(\xi;\eta)$, to the zero-order EP, $V_{00}(\xi)$, in eq. (8a). For this particular realisation, the EP in eq. (7), $V_{eff}(\xi;\eta) = V_{00}(\xi)$, does not depend on $\eta$, and one obtains quasi-free, uniform-density solutions for $\psi_0(\xi)$ which, in their turn, give indeed vanishingly small magnitudes of $\hat{V}(\xi;\eta)$ upon integration in eqs. (8a,b) (since it is reduced to the scalar product of orthogonal quasi-free eigenfunctions, whereas the denominators in eq. (8b) remain always finite). The corresponding eigenvalues of the intermediate realisation, $\eta_{0i}$, form a particular, reduced set of $N_g$ members found above in the analysis of eq. (12) (as opposed to $N_e(N_g)^2$ eigenvalues for each regular realisation), which is explained by the effective absence, in this exceptional case, of the nonlinear EP dependence on $\eta$ and related, physically, to the practically "separated", non-entangled interaction participants (protofields) in this transient, quasi-free configuration.

In that way the intermediate (or "main") realisation of the system with interaction reveals the causal meaning of the canonical, perturbative solution and analysis of the same system: the extended version of "perturbative" solution/realisation exists and describes precisely the quasi-free intermediate system state-realisation during its unceasing transitions between the "regular", highly localised and essentially nonperturbative, internally entangled ("nonseparable") realisations which are equally real and thus taken (in a causally random order) by the system, always passing between them through the "connecting", intermediate realisation. The major deficiency of the perturbative analysis is that, getting into the trap of external "self-consistency" of the "main" realisation (expressing its *relative*, local independence, or "completeness"), it totally ignores all the "regular" system realisations, representing just the most pronounced, *dynamically entangled* result of the whole interaction process as such (see also below), and takes into account a simplified version of only one, very special realisation, representing in reality only the dynamically averaged density of (highly irregular) transitions between the "essential", always dynamically redundant results of interaction. It is clear, in particular, that the latter correspond to the "corpuscular", dynamically localised state of the thus emerging elementary field-particle, whereas the intermediate realisation provides the causal, realistic version of the (now) dynamically related and probabilistic "wavefunction" (see sections 3, 5 for further details), which reveals also the fundamental, irreducible origin of the postulated "quantum mysteries" in the canonical, dynamically single-valued (unitary) theory. We can say also that any version of the canonical "perturbation expansion" always diverges because the system possesses as if the "singularity" of dynamical "branch point" (giving many incompatible realisations) *everywhere* within its causally complete evolution.

The dynamic redundance phenomenon revealed above means that the genuine (i. e. really complete) *general solution* of a problem, describing the unreduced, complex-dynamical interaction process, is obtained at the level of observable system property, generalised density $\rho(q,\xi)$, as a *causally probabilistic sum* of densities, $\rho_r(q,\xi)$, for individual (regular) system realisations:

$$\rho(\xi,Q) = \sum_{r=1}^{N_\Re} {}^\oplus \rho_r(\xi,Q) \, , \tag{15}$$

where $N_\Re$ ($= N_g$ in our case) is the total number of system realisations, and the sign $\oplus$ serves to designate the special, *dynamically probabilistic* meaning of the sum. The latter implies that the *causally derived* densities of regular compound-system realisations, $\rho_r(q,\xi)$ (see eq. (13)), appear and disappear in the actual observations in a *dynamically random*, "spontaneous", "unpredictable", or *chaotic* order, *whatever* is the time of observation and the number of registered events (where *event* is *rigorously and universally defined* now as the totally dynamic realisation emergence and disappearance).

Since the elementary realisations dynamically emerge on absolutely *equal* grounds, we obtain the *a priory determined*, causal (dynamic) probability, $\alpha_r$, of the *r*-th elementary realisation occurrence as

$$\alpha_r = \frac{1}{N_\Re} \quad (r = 1,...,N_\Re), \qquad \sum_{r=1}^{N_\Re} \alpha_r = 1 \, . \tag{16a}$$

As in many cases the elementary realisations are not individually resolved in actual measurements and, being inhomogeneously distributed, appear in dense groups, or dynamic "tendencies", containing various numbers of them, in the general case the probabilities of appearance of such "compound" realisations are not equal, but still *dynamically* determined by the corresponding numbers of constituent elementary realisations:

$$\alpha_r(N_r) = \frac{N_r}{N_\Re} \quad \left(N_r = 1,...,N_\Re; \sum_r N_r = N_\Re \right), \qquad \sum_r \alpha_r = 1 \, , \tag{16b}$$

where $N_r$ is the number of elementary realisations in their $r$-th group, forming an actually observed, compound realisation.

When the number of events (observation time) is large enough, one measures the familiar *expectation (average) value*,

$$\rho_{\text{ex}}(q,\xi) = \sum_{r=1}^{N_\Re} \alpha_r \rho_r(q,\xi) . \tag{17}$$

However, the dynamically probabilistic sum of eq. (15) and the associated causally deduced probability values, eqs. (16), preserve their meaning also for *single, isolated* events emerging in the *real time* of observation/measurement (and thus *together* with the causally specified time itself, see below).

Note that the causally complete picture of dynamically redundant evolution of any "quantum" system means that it is basically *nonunitary*, which can be physically (and mathematically) specified now as its particular structure in the form of qualitatively nonuniform ("jump-like") and dynamically chaotic sequence of discrete events. This intrinsic nonunitarity can be hidden in the canonical "quantum postulates" of the standard, formally unitary description because the individual inhomogeneous, chaotic "steps" of the quantum beat dynamics, forming the very *first* "level of complexity" (dynamical structure) of the world, cannot be resolved in detail in principle, which may leave an impression of a "smooth", unitary evolution within the single-valued dynamics of a unique, "averaged" system realisation (unitarity is thus *equivalent* to single-valuedness and *essential* linearity). However, the underlying real, nonunitary basis of the system dynamics inevitably leaves its "harmful" traces on the artificially smoothened façade of the canonical theory in the form of persistent "unsolvable" problems and "mysteries" in the observed quantum behaviour, like the origin of unceasingly flowing, irreversible time, irreducible unpredictability and discreteness, and various manifestations of duality (or "complementarity").

## 3. Complex-dynamical structure of the electron: quantum beat duality, causal wavefunction, fractal entanglement, and spin

As shown above, the interaction partners (protofields) temporally return, during each system jump between two successive regular (localised) realisations, to a quasi-free state qualitatively close to their (hypothetical) "primordial" state of "ether" before the beginning of the interaction process. Since the reduction centres form a causally random sequence (because of the dynamic redundance phenomenon), it follows that the intermediate realisation is a "uniformly chaotic", rather then "smoothly homogeneous" state. As shown in more detail below (see also [14,16]), this highly irregular, unstable intermediate state of a system with interaction, dynamically "connecting" all regular realisations into a single structure-process, provides the causal, physically real extension of the canonical *wavefunction*, only formally introduced in the conventional quantum mechanics, but now endowed with the realistic, physically complete interpretation and generalised to any interaction process. It is easy to see that the contradictory, mysteriously probabilistic and dualistic wavefunction properties, killing any attempt of its realistic interpretation within any dynamically single-valued theory, are due to the causally probabilistic structure of the intermediate realisation and its permanent dynamic transformation into the regular, "normal" system realisations (corpuscular states of the field-particle).

In particular, the intermediate realisation plays the role of a *dynamically* emerging "surface" (or "time moment") of boundary (or initial) conditions, and the elementary matching of the state-functions for regular realisations, eqs. (10)-(11), or (13), to that of the quasi-free state of the intermediate realisation (wavefunction) confirms the above expressions for regular realisation probabilities $\alpha_r$, eqs. (16), where the realisation numbers are proportional respectively to local ($N_r$) and whole-space ($N_\Re$) integrals of the observed system density (wavefunction squared modulus), which can be understood as the causal substantiation and extension of the canonical "Born's probability rule" [14-16]. This postulate of the canonical theory, contributing to its unsolvable "puzzles", becomes a physically transparent consequence of the complex quantum beat dynamics following from the above equations: the wave field of the intermediate realisation, possessing a delocalised structure that necessarily spans all possible localised, regular system realisations it directly produces, will evidently collapse to a corpuscular state with higher probability there where it has higher (average) density and vice versa, while all its local density variations can only be produced at each other's expense and should therefore mutually compensate, which is expressed as wavefunction squared modulus "normalisation to probability 1".

We see that the protofield interaction process (but also any unreduced interaction at a higher level of world dynamics) has the *dynamically complex* internal structure (in a well-specified sense), where the phases of *strong* interaction and physical entanglement within each regular realisation *alternate* with transitional periods of *weak* interaction of quasi-free partners within the intermediate realisation, without any artificial insertion of this nonlinear pulsation in the initial problem formulation. The actually perceived "magnitude" of interaction is usually the natural, dynamical average between all realisations and especially those two qualitatively different, always present components of "strong" and "weak" interaction. In terms of the above causally extended wavefunction, this intrinsic duality of complex dynamics takes the form of "mysterious" (and now completely understood) combination of the "quasi-free" and therefore basically "linear" state of the wavefunction (obeying, in particular, the *derived* linear Schrödinger equation [14]) with its highly nonlinear behaviour (though hidden in the canonical postulates) as far as "measurement", corpuscular, and "quantum-transition" properties of the same objects are involved (cf. ref. [9]).

In that way we have already revealed three closely related aspects of the same, now causally explained, property of "wave-particle duality": (i) the unceasing, physically real, and totally internal transformation between the well-defined squeezed and delocalised states of the elementary field-particle, (ii) the corresponding alternation of the "strong" ("nonperturbative") and "weak" (quasi-"perturbative") interaction regimes in the underlying coupled protofield system, and (iii) combination of the (alternating) "linear" and "nonlinear" types of the field-particle behaviour. We see now that all of them are but particular manifestations of the intrinsic duality of complex dynamics of *any* unreduced interaction process, explaining the fundamental meaning of the conventional Bohr's "complementarity", which is intuitively "felt" as a more general basis behind the observed particular manifestations of quantum weirdness.

The described *unreduced dynamic complexity* of the natural, dynamically multivalued interaction processes (quantum beat for the coupled protofield system) can be further specified within a universally applicable and totally consistent measure of complexity by any growing function of the number of system realisations (or the rate of their change) equal to zero for only one realisation [14] (see ref. [16] and below for further details). Being expressed by various measurable quantities, such as energy [14,16], thus defined dynamic complexity is directly related to the above properties of dynamic redundance and causal randomness, which means that the unreduced complexity is actually a *synonym* of the causally extended notion of *dynamical chaos*, characterising the inseparable mixture of regularity of each individual realisation and dynamic randomness of the order of their natural emergence in the form of physically real space structure.

Another closely related feature of the unreduced complexity of natural interaction processes is the autonomous dynamic *entanglement* of interacting entities within any regular realisation and their transient *disentanglement* during system transitions between realisations (through the phase of intermediate realisation, or causal wavefunction). The physically real dynamic entanglement is described by sums of weighted, "interwoven" products of functions of $q$ and $\xi$ in the expressions, eqs. (10)-(11), (13), for the compound system state-function, $\Psi_r(q,\xi)$. Moreover, the entanglement continues, by the same fundamental mechanism, at ever smaller scales related to the current, explicitly considered scale through the solutions, {$\psi_{ni}^0(\xi)$}, {$\eta_{ni}^0$}, of the "auxiliary" system of equations, eqs. (9). We shall not explicitly provide here the application of the same analysis to that hierarchy of scales described by the corresponding hierarchy of equations and revealing the complex-dynamical, *causally probabilistic* extension of *fractal* (the details can be found in ref. [14]), also because in the case of interacting protofields those smaller scales are directly unobservable from within this world, even though their existence is necessary for the formation of the smallest observable structures, the elementary field-particles. It is important that the fractal, "infinitely fine" dynamic entanglement of the interacting e/m and gravitational "degrees of freedom" constitutes the real, *physically* (and *therefore* also mathematically) "inseparable" structure within the core of the elementary field-particle (like the electron) in its corpuscular, dynamically squeezed state, as it is described by eqs. (13), (14). Another fundamentally important role of the fractal entanglement of interacting entities is that it determines the unavoidable dynamic instability of any realisation, even in the total absence of additional, "noisy" perturbations, so that each localised, corpuscular phase (realisation) of the field-particle should necessary be transformed into its extended (undulatory) state in the intermediate realisation, and vice versa. The process of disentanglement, or extension, of interaction partners up to their transient quasi-free state in the intermediate realisation is opposite to the dynamic entanglement, or reduction. It has already been described above and corresponds to temporary quasi-perturbative separation of the effective existence equation, eq. (7), for that particular realisation.

Note that the revealed dynamic redundance and entanglement properties actually form two inseparable aspects of the same unreduced interaction complexity: redundance of the interaction results means that the unreduced entanglement/combination of interacting entities has the redundant number of equally real versions (see eq. (12) and its discussion above), which leads to their unceasing causally random change within the emerging quantum beat process. Correspondingly, none of the related manifestations of the unreduced dynamic complexity can be obtained within its single-valued conventional imitations (see ref. [14] for more details), always using perturbation expansions that inevitably kill, as we have seen, the essential links in the natural interaction development (which does not prevent one from a purely empirical, or numerical, observation/description of the resulting properties that remain, however, unexplained in their fundamental origin). Therefore the true reason for the persistently "mysterious" character of "quantum-mechanical" duality, uncertainty and other related properties within the conventional approaches, including their latest, always axiomatic "interpretations" (like e. g. "quantum histories"), is their invariably single-valued, perturbative, "anti-interactional" and purely abstract character neglecting the unreduced dynamic complexity of real interactions and its universal, irreducible, and therefore omnipresent manifestations (dynamic randomness, entanglement, duality, etc.). In particular, the formal, typically mysterious "quantum entanglement" of the abstract "state vectors" often evoked in the conventional quantum mechanics, and especially recently in relation to "quantum computers", is none other than the unitary, mechanistically simplified imitation of the physically real, complex-dynamical entanglement described above, which explains the persistence of the mystery and the related fundamental deficiency of the concept of unitary quantum computation (see ref. [14] for more details and references).

We cannot avoid mentioning here another irreducible feature of the complex reduction-extension dynamics of the interacting protofields giving rise to the causal understanding of the universal property of *spin* of the emerging elementary particle and providing further refinement of the fractal dynamic entanglement process (a more detailed analysis is proposed in ref. [14]). The above picture of the dynamical, physically real interlacement of the protofields in the process of squeeze to a very small volume implies already a clear image of the necessary rotational component of motion of the fields around each reduction centre. This tentative image can be provided with a fundamental substantiation if we note that the self-sustained squeeze of an extended protofield with a finite compressibility to a point-like state should inevitably result in the shear instability within the field leading to its vortex-like motion towards the centre of reduction. The phenomenon is generally similar to the (turbulent) forced passage of a liquid through a small hole, but in our case we have its much more "nonlinear", self-amplified version, while the unique role of the complex (unreduced) field-particle dynamics consists in "automatic", unceasing creation of a sequence of those "small holes" (any canonical, unitary theory would predict for this case a single, more or less homogeneous "fall" of one of the protofields onto another). It is this highly nonuniform and internally irregular vortex motion of the elementary field-particle in course of each reduction (and the following extension) which provides both causal, universal understanding of the property of spin and further refinement of the process of dynamic entanglement of the interacting (physical) protofields, giving the adequate picture of the dynamic "internal structure", and thus *physical* essence, of the *elementary particle* as such, the electron being its simplest, "canonical" species. Note, however, that as the EP magnitude can be estimated by the particle mass, the relatively light particles, such as electron and other leptons, are characterised by a weak degree of direct protofield "mixture", remaining "wandering" nonlinear vortices, or *virtual solitons*, of the e/m protofield driven by attraction to the hidden gravitational protofield, whereas the heavier, hadronic particles correspond to higher EP magnitudes and realise a much more close, direct protofield entanglement (see refs. [14,16] for more details).

## 4. Causal relativity of complex elementary particle dynamics

Consider the elementary field-particle, such as electron, remaining globally at rest. According to the previous analysis (sections 2, 3) [14-16], it is represented by the unceasing, intrinsically random process of quantum beat in the system of coupled e/m and gravitational protofields/media, equivalent to spatially chaotic wandering of the emerging virtual soliton. We can state now that this unceasing chaotic motion appears in observations as the *universally* determined property of *inertia* of the elementary particle (and eventually, of macroscopic bodies composed from such particles) measured by the field-particle *mass*, which is *causally equivalent* to the rest *energy* appearing in the form of chaotic wandering of the virtual soliton (or system realisation change). Indeed, it is clear that any attempt to change the dynamical state of the field-particle will meet an effective resistance due to the *already existing* internal, "thermal" (chaotic) motion of the virtual soliton, which is generally similar to resistance to compression of a gas consisting in this case of only one "molecule" that possesses, however, perfectly chaotic dynamics and therefore tries to "take all the accessible volume" (of system realisations). Such definition of inertial mass as an *emergent*, dynamical property is possible only due to the causally derived self-sustained character of the quantum beat process and its *purely dynamic* chaoticity, obtained in the form of dynamic multivaluedness (redundance) of the protofield interaction process.

In that way, we are able to extend the concept of the "variable mass" of a particle in de Broglie's "hidden thermodynamics" [10] because we can replace *additional* chaotic perturbations of the regular, "inherent" particle motion and mass coming from the external "hidden thermostat" ("subquantum medium") in the original de Broglie's picture by the intrinsic, purely dynamic chaoticity emerging in the process of *a priory* totally regular (and closed) protofield interaction system and determining the *whole* field-particle dynamics. The same refers to other reappearing attempts to describe inertia as a result of formally introduced *external* influences, such as "vacuum fluctuations" [20] or omnipresent field/particles ("Higgs bosons"), on an initially massless "particle" or "charge" of a postulated, but poorly specified physical origin, which is actually equivalent to a considerably simplified, but externally, terminologically "updated" version of the original de Broglie's picture [9-10].

The proposed universal, physically transparent and fundamentally substantiated interpretation of inertia/energy should now be provided with an equally consistent mathematical expression. In order to properly account for the unreduced complexity of the quantum beat dynamics, the intrinsic mass-energy of the elementary particle-process described above should be expressed through the unique and universal measure of that complexity naturally reflecting also the major property of discreteness of quantum behaviour and integrating its emerging spatial and temporal aspects.

The spatial structure element of the electron is the length, $\Delta x$, of the system jump between two localised states determined by the eigenvalue spacing between the neighbouring realisations, $\Delta \eta_i^r$, found from the effective existence equation, eqs. (7)-(9). The physical time element, $\Delta t$, emerges as the quantum beat period, related to $\Delta x$ by the usual condition $\Delta t = \Delta x / c$ (where the speed of light, $c$, is *defined* as the speed of perturbation propagation in the e/m protofield, coupled to the gravitational medium). The simplest, universally defined physical quantity, which is independently proportional to $\Delta x$ and $\Delta t$, is the *action* increment in one quantum beat cycle, $\Delta \mathcal{A} = -E\Delta t + p\Delta x$, where $E$ and $p$ are coefficients to be eventually identified with energy and momentum by analogy with classical mechanics for which our universal description should remain valid. Therefore we can suppose that it is the mechanical action, $\mathcal{A}$, which provides, within its *extended*, causally complete understanding, the desired universal complexity measure. It corresponds to the above general definition of complexity through the total realisation number (section 3) because $\Delta \mathcal{A}$ describes an elementary realisation change and its integration gives a quantity proportional to system realisation number. But since quantum level of world dynamics is characterised by the universal quantum of action in the form of Planck's constant $h$, it is clear that for processes with elementary particles $|\Delta \mathcal{A}| = h$ is the universal quantum of complexity, which provides a new, deeper understanding of the physical meaning and importance of $h$ [14,21]. It means that each reduction-extension cycle of the *essentially nonlinear, intrinsically unstable, dynamically discrete* quantum beat process (sections 2, 3) corresponds to a change of $h$ in the action value attributed to the system, $|\Delta \mathcal{A}| = h$, which involves a considerable extension of the notion of action itself with respect to the one from the essentially linear, unitary mechanics with its uniform evolution and smooth trajectory/path of the system motion.

Recalling the canonical relation between energy, action, and time in the classical mechanics, $E = -\partial \mathcal{A}/\partial t$, that should remain valid within the realistic picture of quantum mechanics, necessarily giving rise to causally emerging classical behaviour (see section 7), we can summarise the obtained results within the following fundamental, *uniquely* determined expression for the introduced field-particle rest energy, $E_0$, representing it as the *temporal rate* (or "intensity") of the underlying *complex-dynamical (spatially* *chaotic)* process of quantum beat:

$$E_0 = -\frac{\Delta \mathcal{A}}{\Delta t} = \frac{h}{\tau_0} = h\nu_0 , \qquad (18)$$

where $-\Delta \mathcal{A} = h$ is the quantum of action-complexity corresponding *physically* to one cycle of the quantum beat process (essentially nonlinear, avalanche-type reduction-extension of the coupled protofields or one "quantum jump" of the "virtual soliton"), $\Delta t = \tau_0$ is the *emerging* "quantum of time" equal to one period, $\tau_0$, of the same cycle, determined by each two consecutive, *real events* of autonomously produced (and consistently derived above) dynamical squeeze (reduction), and $\nu_0 \equiv 1/\tau_0$ is the quantum beat *frequency*, forming thus the basis of the causal time concept [14,16] ($\nu_0 \sim 10^{20}$ Hz for the electron).

Since according to the above definition energy is the measure of intensity, or inertia, of the internal chaotic dynamics of a particle, it should be proportional to particle mass, $E_0 = m_0 c^2$, where for the moment $c^2$ is simply a coefficient (we shall properly specify this relation later). This permits us to present the above expression for the rest energy, eq. (1), in an equivalent form,

$$m_0 c^2 = h\nu_0 , \qquad (19)$$

which was semi-axiomatically introduced by de Broglie and further used in his original derivation of the wave properties of elementary particles, culminating in the famous expression for the "de Broglie wavelength" [6]. We emphasize that in our version the simple relation of eq. (19) is causally *derived* from the physically based consideration of the underlying complex dynamics of interacting protofields and represents just a compact expression of the essentially nonlinear, fractally entangled, dynamically redundant (and thus probabilistic) quantum beat process (in particular, the fundamental frequency $\nu_0$ represents "something periodic", but actually much more involved than any linear, or even ordinary "nonlinear", oscillation). This is the beginning of the natural appearance, within our first-principles analysis, of the *causally extended, physically based relativity* which is *intrinsically unified* with the quantized wave dynamics of the elementary field-particle and its dualistic internal structure, being simply *another manifestation* of the *same* dynamic complexity that underlies quantum behaviour of the system of coupled protofields [14,16].

If now the isolation of the elementary field-particle is violated and it is subjected to external influences (coming from other particles and transmitted through the common physical "world manifold" of two coupled protofields), then the intensity of the quantum beat, and thus particle energy, can change. We can rigorously define the *state of rest* of the elementary field-particle (and eventually of a macroscopic body consisting from such particles) as the state with *minimum* complexity-energy. This minimum exists, since the quantum beat energy is always positive and finite. For the elementary field-particle it is realised as *totally irregular* spatial distribution of the reduction centres, which corresponds to the absolutely homogeneous distribution of realisation probabilities, causally deduced above within the generalised Born's rule (section 3) or simply as a result of "equal rights" for emergence of apparently equivalent realisations (eqs. (16a) in section 2).

It is not difficult to make the straightforward next step and introduce an equally rigorous and universal definition of a *state of motion* as any system state characterised by a complexity-energy value *exceeding* the minimum value of the state of rest for this system. Naturally, there can be many such higher values of energy and many respective states of motion. Since the particle state of rest has a unique, totally uniform spatial structure, it is clear that any state of motion of the field-particle will be characterised by an inhomogeneous spatial distribution of probabilities of reduction to different centres (system realisations), corresponding to a *dynamic tendency* in the thus emerging *global* motion of the field-particle, always remaining a well-defined *complex-dynamical process* [14,16]. Similarly, any state of motion of a more complex, necessarily inhomogeneous, object is characterised by a *more* inhomogeneous distribution of realisation probabilities than that for its state of rest. Whereas the centre of each next reduction event is "chosen" by the system always in a purely probabilistic fashion, there is now more "order in chaos", and the emerging more inhomogeneous structure in the reduction (realisation) *probability distribution* manifests itself as observable *spatial degrees of freedom* (or "spatial structure") of the progressing "(sub)level of complexity" of the system, that is the particle/object "displacement" as such. In terms of the proposed mathematical measures of complexity (action and energy-mass), this means that our action-complexity $\mathcal{A}$ for the field-particle in a state of motion acquires a (regular) *spatial dependence* (structure), whereas in the state of rest it depends only on time. Therefore the partial time derivative of action of the moving field-particle, always defining its energy, is now different from the total time derivative, in full agreement with the well established relations from the classical mechanics:

$$\frac{d\mathcal{A}}{dt} = \frac{\partial \mathcal{A}}{\partial t} + \frac{\partial \mathcal{A}}{\partial x}\frac{dx}{dt} = -E + pv, \tag{20}$$

where the *momentum*, $p$, that characterises the emerging spatial order in the probability distribution of the moving field-particle and its *velocity*, $v$, are introduced in accord with the canonical relations:

$$p = \frac{\partial \mathcal{A}}{\partial x}, \qquad v = \frac{dx}{dt}.$$

Now we should take into account the natural discreteness (dynamic quantization) of the quantum beat process meaning, in particular, that the spatial field-particle structure can emerge only as discrete elements with spatial *dimension* (size) $\lambda$ determined always by the same quantum of action, $h$:

$$p = \frac{\Delta \mathcal{A}}{\Delta x}\bigg|_{t=\text{const}} = \frac{\Delta \mathcal{A}}{\lambda} = \frac{h}{\lambda}, \tag{21}$$

$$v = \frac{\Delta x}{\Delta t} \equiv \frac{\Lambda}{T}, \tag{22}$$

where $\lambda \equiv (\Delta x)|_{t=\text{const}}$ is the *emerging* "quantum of space" of the current (actually lowest observable) level of complexity, a minimum directly measurable (regular) space inhomogeneity characterising the elementary quantum field with complexity-energy $E \,(> E_0)$ and resulting from its global displacement (motion), $\Delta t = T$ is the "total" period of nonlinear quantum beat of the field-particle in the state of motion with complexity-energy $E$ ($N = 1/T$ is the corresponding quantum-beat frequency), $\Delta x = \Lambda$ is the "total" quantum of space, while $\tau \equiv (\Delta t)|_{x=\text{const}}$ is the quantum-beat oscillation period measured at a fixed space point (so that $E = h/\tau$). We can therefore rewrite eq. (20) in the following form, specified for the quantum beat dynamics of the moving field-particle:

$$E = -\frac{\Delta \mathcal{A}}{\Delta t} + \frac{\Delta \mathcal{A}}{\lambda}\frac{\Delta x}{\Delta t} = \frac{h}{T} + \frac{h}{\lambda}v = hN + pv, \tag{23}$$

where

$$E = -\frac{\Delta \mathcal{A}}{\Delta t}\bigg|_{x=\text{const}} = \frac{h}{\tau} = h\nu \ . \tag{24}$$

The relation of eq. (23) replaces its limiting case of eqs. (18), (19) valid for the same field-particle at rest, and includes the rigorously specified and causally substantiated definitions of the emerging space, time, momentum, and energy, applicable also in the general case. The causally defined field-particle spin (section 3) can be taken into account within the same relation [14], but in order to emphasize most important new moments, we shall not do it explicitly here, limiting ourselves to the case of negligible spin-related effects. Note that the characteristic temporal intervals for the moving particle, $\tau$ and $T$, are both different from the single time scale $\tau_0$ for the particle at rest because the modified intensity, and thus temporal rate, of the quantum beat process constitutes the real physical mechanism of the global motion as such: in order to advance as a whole, the particle should *relatively* (with respect to the state of rest) intensify (make more frequent) its reduction-extension cycles "in the direction of motion" at the expense of those in other directions, which reveals already the fundamental *physical* source of the "relativity of time" (it will be further specified below). We see also that the appearance of global motion is naturally involved with emergence of an elementary spatial structure of the moving field-particle, in the form of an average regular tendency within the generally chaotic wave field. It is not difficult to understand that this averaged structure is none other than the causally extended, realistic "de Broglie wave of the (moving) particle", whereas its characteristic size, $\lambda \equiv \lambda_B$, is the "de Broglie wavelength", and we shall continue to confirm and specify this conclusion below. The physical meaning of the energy partition of eq. (23) becomes clear: the second term, $pv = h\nu/\lambda$, corresponds to the regular, global motion tendency, or structure, whereas the first term, $-(d\mathcal{A}/dt) = h/T$, represents purely irregular, "thermal" wandering of the virtual soliton "around" the average tendency/structure. This dynamically based interpretation confirms and completes the corresponding "thermodynamical", phenomenological considerations of Louis de Broglie [10].

Further refinement of mathematical description of the quantum beat dynamics for the moving field-particle, eq. (23), should produce a relation between the temporal ($E$) and spatial ($p$) rates of the complex-dynamical structure formation process called, in accord with tradition, *dispersion relation*. The holistic character of the quantum beat dynamics shows that such a relation should exist in a well-defined form and can be rigorously deduced: the regular global structure of the wave field ($p$) is formed by the same reduction-extension cycles as those determining the general quantum beat intensity ($E$). The dispersion relation we are looking for can be considered as the causally completed version of the "phase accord theorem" introduced by Louis de Broglie in his original substantiation of existence of a wave associated with a particle [6]. While the wave existence is *assumed* and its detailed physical origin remains unclear within this original formulation, de Broglie shows that if the "internal oscillations" of the particle, also *postulated* by the heuristically composed analogue of eq. (19), remain always in phase with those of the wave (i. e. physically the wave performs the stationary transport of, "pilots", or "develops", the particle-oscillation), then such "compound object" moves in accord with the relativistic transformations of time and mass (also *formally postulated* in the standard relativity) and the length of the wave is given by the (now) canonical expression for the "de Broglie wavelength". The quantum beat dynamics involves a self-consistently unified, causal extension of the participating entities and their "phase accord": the localised "particle" (virtual soliton) moves always "in phase" with the extended wave ("intermediate realisation of the system", section 3) simply because it is permanently *transformed* into that wave and back, in course of its "oscillations" (represented in reality by successive, essentially nonlinear reductions and extensions of the coupled protofields). In other words, the internal "particle oscillations" and the "piloting" wave motion represent two dynamically related, *alternating* and *therefore* coexisting aspects of *one and the same* quantum beat process. Another essential extension of the original phase-accord idea is in the fact that the regular, averaged structure of the entire wave field, forming the causal analogue of the "transporting wave", represents only a part of the whole wave-field dynamics, and the localised virtual soliton, or "particle", makes many essential, irregular *deviations* from that regular tendency, "taking" with it the corresponding extended wave structures that form in this case a causally random, irregular part of the wave field, always preserving, however, the internal "phase accord" between "particle" and "wave". It is clear that only the averaged regular structure of the wave field, the causally completed version of the "de Broglie wave", can be directly observed in experiments as a wave, while the stochastic, purely random components of the field-particle dynamics appear only "statistically", in the form of the particle (rest) mass.

Consider a portion of that regular undulatory spatial structure of the wave field in the state of global motion incorporating $n$ (de Broglian) wavelengths, $x_0 = n\lambda$. The temporal, oscillatory aspect of the quantum beat process determining the (total) energy covers the (measured) distance, $x_0$, *in the direction of motion*, remaining in the described involvement with the wave, but simultaneously performs extensive irregular, "sideways" deviations (wandering). Since the velocity of the virtual soliton propagation along the actual irregular path should always be equal to the velocity of light, $c$ (it is the velocity of the directly observable perturbation propagation in e/m medium/protofield), it becomes clear that the energy-bearing oscillation of the quantum beat actually performs $n' = n(c/v)$ cycles within the observed portion of the regular displacement with velocity $v$. It is physically obvious why always $n' > n$ and thus $v < c$: the difference between $n'$ and $n$ accounts for the "hidden", purely irregular part of the quantum beat process. The measured spatial advance of the oscillation (temporal aspect), $x_1 = n'\tau c = n(c^2/v)\tau$, should be equal to the observed displacement of the wave (spatial aspect), $x_0 = n\lambda$, within the holistic quantum beat process: $x_0 = x_1$, or $\lambda = V_{ph}\tau$, where $V_{ph} = c^2/v$ is the fictitious superluminal "phase velocity" introduced by de Broglie within the original phase accord conjecture [6]. Rewriting the obtained relation between $\lambda$ and $\tau$ as

$$\frac{1}{\lambda} = \frac{1}{\tau}\frac{v}{c^2},$$

multiplying it by $h$, and using the definitions for momentum and energy, eqs. (21), (24), we finally obtain the desired dispersion relation between spatial (regular) and temporal (total, regular and irregular) aspects of complex field-particle dynamics:

$$p = E\frac{v}{c^2} = mv, \qquad (25)$$

where $m \equiv E/c^2$, now by *rigorously substantiated* definition in which $c$ is the velocity of light. We have thus causally deduced the famous "relativistic" dispersion relation from the detailed analysis of the unreduced *complex (chaotic) dynamics* actually underlying any externally "uniform" motion (whereas it is derived from formal *postulates* in the canonical theory, see e. g. refs. [22,23]) and simultaneously specified the law of proportionality and the coefficient, $c^2$, in the yet more famous energy-mass relation, already physically substantiated above and also only mechanistically "guessed" in the standard relativity.

We can see now the origin of the characteristic inconsistency around the superluminal de Broglian "wave of phase" [6], being both physically unreal and indispensable for the proposed realistic approach: if one does not explicitly take into account the really existing, intrinsically irregular part of the holistic, inseparable quantum beat dynamics, then the corresponding part of energy is "forced" to reappear in a reduced form of regular motion, inevitably becoming "superluminal" because of its "excessive" quantity. The problem solution proposed by de Broglie and transforming the ambiguous "wave of phase", propagating with the velocity $V_{\text{ph}} = c^2/v$, into a real wave transport of the particle by the linear *wave packet* moving with the proper *group velocity* $v$, cannot really solve the problem, since the purely *linear* transformation between the phase and group wave propagation shows fundamental divergence with the necessarily and essentially *nonlinear* character of the wave mechanics (in particular, the linear wave packets quickly spread etc.). The realistic undulatory behaviour of the moving elementary particle should indeed be obtained from *many* participating, slightly *differing* components, but in an adequate, essentially nonlinear description they should necessarily (strongly) *interact* with each other, which leads, as shown above (sections 2, 3), to the intrinsically unstable character of the resulting quantum beat dynamics. The nonlinear component interaction takes the form of their ruthless *competition* in which only those corresponding to the current tendency (reduction to a randomly selected centre or the opposite extension) can survive and form both the localised "particle" and its "wave transport". Only the unreduced involvement of the largely hidden, but really existing complex-dynamical processes including *many incompatible* state-realisations can consistently explain the externally simple "relativistic" and "quantum" relations between mass, energy, and momentum, eqs. (19), (25), and it is the insertion of the physically substantiated and rigorously derived dispersion relation, eq. (25), into the emerging momentum definition, eq. (21), that finally closes the causally complete substantiation of de Broglie's expression for the wavelength:

$$\lambda = \lambda_{\text{B}} = \frac{h}{mv} \ . \tag{26}$$

We emphasize the crucial importance of the obtained dispersion relation, eq. (25), for the complete understanding of the meaning of eq. (26). It is the deceptively familiar, "classical" and almost "trivial" relation between the particle momentum, mass, and velocity, $p = mv$, that bounds together the wave-like, corpuscular (classical) and relativistic properties of the elementary field-particle into the intrinsically inseparable (within the unreduced description) and most "mysterious" (within the conventional theory) mixture summarised by the fundamental de Broglie equation. Note also that the same dispersion relation is actually equivalent to the main dynamical laws of classical mechanics (*Newton's laws* and their relativistic extension), which are obtained in their canonical form by taking the time derivative of eq. (25). We see now that these laws can also be *causally derived*, instead of being postulated in the canonical, dynamically single-valued approach, if we take into account the internal complex (multivalued) dynamics of *any* motion, including all the externally "uniform" and "rectilinear" cases (the detailed mechanism of transition from quantum to classical behaviour is outlined in section 7). The applicability of these conclusions to any macroscopic, many-particle body follows not only from the fact that they apply to every its part, but from the unlimited effective universality [14] of our description of unreduced interaction dynamics (section 2), playing thus an important role as reflection of the necessarily profound coherence within the multi-level complex-dynamical world construction (we continue to specify this conclusion below, in sections 6, 7).

The emerging causal unification of the (extended) relativity and quantum mechanics within our theory does not stop there, and we continue to specify it by inserting the causally derived dispersion relation, eq. (25), into the basic energy partition of eq. (23) and using the complex-dynamic energy definition, eq. (24), which gives:

$$\tau = T\left(1 - \frac{v^2}{c^2}\right). \tag{27}$$

This relation provides further refinement of the *causally complete* explanation of the "relativistic time retardation" that *cannot* be separated from the unreduced, realistic understanding of the *entity of time* itself (absent in the conventional relativity [22,23]). Since time naturally *emerges together* with structure-forming reduction-extension events that constitute the quantum-beat global *motion*, the increased intensity (rate, frequency) of this process for the (relatively) moving field-particle(s) is realised as a relative decrease of the elementary quantum beat period, $\tau$, which results in *relative* retardation of *any other* process from a higher level of complexity within the moving group of particles (a physical "body"), *irrespective* of their detailed dynamics (their duration with respect to the state of rest is proportional to the relatively increased "internal" quantum beat period, $T$). An equivalent formulation involves "transport effects" for the appearing undulatory structure and is determined by the difference between the total and partial time derivatives of the field-particle action-complexity (see eq. (20)): they lead to a relative increase of the local "total-change" (irregular-motion) period $T$ that just determines relative local duration of any internal motion cycle. In other words, causal time within the moving system, *with respect* to the state of rest, goes as $nT$ (with permanently and *autonomously* growing integer $n$), and the relative duration of any given process equals to $n_0 T$, where $n_0$ depends only on the internal process type, but not on its global state of motion.

In order to obtain the causal time retardation effect in the explicit form, we should now express the quantum beat periods $\tau$ and $T$ through the reference period in the state of rest, $\tau_0$. It is not difficult to see [14] that the corresponding frequencies, $\nu$, $N$, and $\nu_0$ are related by the following equation:

$$N\nu = (\nu_0)^2, \tag{28a}$$

which gives, for the periods,

$$T\tau = (\tau_0)^2. \tag{28b}$$

These relations can be reduced to a physically transparent law of "conservation of the total number of reduction events" which is a manifestation of the universal complexity conservation law [14]. In our case, it stems simply from the fact that the driving electro-gravitational coupling remains unchanged for any global motion state of the field-particle, and therefore the total number of reduction events (per unit of time), including both the regular tendency and irregular deviations, should be the same in the states of rest and (uniform) motion (respectively the right- and left-hand sides of eq. (28a)), while the relative *proportions* (intensity) of global regularity and random deviations from it change between different states of motion (and rest). Using eqs. (28) in conjunction with eq. (27), we get the canonical expression of the time retardation effect, now causally explained by the underlying physical picture of *complex quantum beat dynamics*:

$$T = \frac{\tau_0}{\sqrt{1 - \frac{v^2}{c^2}}} \quad \text{or} \quad N = \nu_0 \sqrt{1 - \frac{v^2}{c^2}}, \tag{29a}$$

$$\tau = \tau_0 \sqrt{1 - \frac{v^2}{c^2}} \quad \text{or} \quad \nu = \frac{\nu_0}{\sqrt{1 - \frac{v^2}{c^2}}}. \tag{29b}$$

We can now combine in one expression the complex-dynamical partition of the total energy into the regular (global) transport and irregular ("thermal") wandering, eq. (23), with the causally derived dispersion relation, eqs. (25), (26), and time (frequency) relation to dynamics, eqs. (29):

$$E = h\nu_0\sqrt{1-\frac{v^2}{c^2}} + \frac{h}{\lambda_B}v = h\nu_0\sqrt{1-\frac{v^2}{c^2}} + h\nu_B = m_0c^2\sqrt{1-\frac{v^2}{c^2}} + \frac{m_0v^2}{\sqrt{1-\frac{v^2}{c^2}}} \,, \qquad (30)$$

where $h\nu_0 = m_0c^2$ (eq. (19)) and we introduce *de Broglie frequency*, $\nu_B$, defined as

$$\nu_B = \frac{v}{\lambda_B} = \frac{pv}{h} = \frac{\nu_{B0}}{\sqrt{1-\frac{v^2}{c^2}}} = \nu\frac{v^2}{c^2} \,, \quad \nu_{B0} = \frac{m_0v^2}{h} = \nu_0\frac{v^2}{c^2} = \frac{v}{\lambda_{B0}} \,, \quad \lambda_{B0} = \frac{h}{m_0v} \,. \qquad (31)$$

Note that the causal relativity of spatial dimensions (of an object), being less important for the main subject of this paper, is obtained within the same type of analysis and enters, in a specific form, to the causally derived expression for the elementary observable length, the de Broglie wavelength, actually coinciding with the above eq. (31):

$$\lambda_B = \lambda_{B0}\sqrt{1-\frac{v^2}{c^2}} = \frac{h}{m_0v}\sqrt{1-\frac{v^2}{c^2}} \,. \qquad (31´)$$

The summarised expression of the complex dynamics of a moving field-particle, eqs. (30), (31), including the "ordinary" type of relation between the causal de Broglie wavelength, frequency, and wave/particle velocity, $\lambda_B\nu_B = v$, clearly demonstrates the physical reality of this wave and demystifies its origin. In particular, $\nu_B$ is the *average* frequency of the dynamically quantized, step-like propagation of the "travelling" de Broglie wave that advances to $\lambda_B$ during each indivisible and irregularly occurring step (coinciding with the corresponding virtual soliton jump). In this respect the averaged, de Broglian wave field of a moving field-particle resembles, due to the internal nonlinearity, both travelling and standing undular structure, the latter having spatially fixed nodal planes. This averaged, global-motion tendency, represented by the second summand in eqs. (30), could not exist without purely irregular deviations from it ("thermal motion"), represented by the first term in eqs. (30), and their ratio, $R$, equal to the ratio of the corresponding frequencies, shows the relative number of probabilistic quantum jumps falling within the regular tendency and taken with respect to the number of purely irregular events:

$$E = m_0c^2\sqrt{1-\frac{v^2}{c^2}}(1+R) \,, \quad R \equiv R(v) = \frac{v^2/c^2}{1-\frac{v^2}{c^2}} = \frac{\beta^2}{1-\beta^2} \,, \quad \beta = \frac{v}{c} \,. \qquad (30´)$$

It is clear that $\alpha_1 = \beta^2 = v^2/c^2$ and $\alpha_2 = 1 - \alpha_1 = 1 - \beta^2 = 1 - v^2/c^2$ are the dynamically defined probabilities that an elementary reduction-extension event falls within, respectively, the regular (averaged, global) and purely irregular ("thermal") tendency of the complex quantum-beat dynamics, while $R = \alpha_1/\alpha_2 = \alpha_1/1-\alpha_1$ (cf. the general definition of such "compound realisation" probabilities $\alpha_r$, eq. (16b) in section 2). We see that a slow, "nonrelativistic" motion ($v \ll c, \beta \ll 1$) is characterised by a very weak "order in chaos", $R, \alpha_1 \ll 1, \alpha_2 \cong 1$, which means that the distribution of realisation (reduction) probabilities is only slightly inhomogeneous and only a small portion of the probabilistic "quantum jumps" of the system and the related wave-field structures falls within the regular tendency forming de Broglie wave. The structure of the probabilistic wave field becomes remarkably ordered, $R \sim 1, \alpha_1, \alpha_2 < 1$, only at moderate relativistic velocities of the global motion ($v \sim c, \beta \sim 1$). And finally, it is almost totally (but *never* completely) ordered, $R \gg 1, \alpha_1 \cong 1, \alpha_2 \ll 1$, at "ultra-relativistic" velocities ($v \cong c, 1 - \beta \ll 1$). The relativistic ordering of the internal soliton wandering around the regular tendency explains, in particular, why the high-energy "quantum" particles show little observable deviations from the classical, "trajectorial" type of behaviour, even though nothing points to this feature in the canonical quantum postulates basically separated from any "relativistic effects", in the canonical interpretation.

Thus the causally complete, dynamic meaning of "relativistic" motion and velocity is revealed within the unreduced picture of complex dynamics of the elementary field-particle. The "relativity" of the observed temporal and spatial aspects of dynamics is an *explicit, direct* manifestation of its internal *complexity* (dynamic multivaluedness), fundamentally inseparable from "quantum" manifestations of the *same* dynamic complexity (quantized spatial and probabilistic temporal structure of the protofield interaction process), even though the two types of manifestations *seem* not to be directly related, both in practical observations and within the canonical, unitary versions of quantum mechanics and relativity, *both* of them actually describing, as we can clearly see now, only the external, effectively one-dimensional "shape", or projection, of the real, dynamically multivalued motion. The increasingly relativistic character of dynamics with the growing global motion velocity corresponds to growing dynamic complexity, measured by such related quantities as *E*, *p*, *R*, and *β*. Physically this change corresponds to increasing order within the causally chaotic dynamics, always preserving, however, its intrinsically probabilistic character at every single step (quantum jump of the virtual soliton). This latter feature explains why any dynamically complex, *massive* particle cannot not only exceed, but even attain the velocity of light: the condition $v = c$ ($\beta = 1$) would correspond to the totally regular, zero-complexity dynamics, where the field-particle could not have any possibility for irregular deviations from the global tendency, while the latter would lose the intrinsically probabilistic character of *each* virtual soliton jump. Only massless, totally regular perturbations of the e/m protofield (like photons [14]) propagate with the speed of light.

We can *physically* understand now why the global, *regular* motion of a massive field-particle (and thus of any "ordinary", structure-forming matter) necessarily involves the "relativistic" increase of its total *inertial* mass, i. e. why *any* energy possesses the property of inertia: this is because *any* part of the total complex process of motion, including the regular *in average* global tendency, occurs through an intrinsically *random* choice of each participating individual reduction centre, thus giving rise to inertia, according to the above causal interpretation of the latter. Therefore, as follows from eqs. (30´),

$$m = \frac{E}{c^2} = m_0 \sqrt{1 - \frac{v^2}{c^2}} \left( 1 + \frac{v^2/c^2}{1 - \frac{v^2}{c^2}} \right) = \frac{m_0}{\sqrt{1 - \frac{v^2}{c^2}}} \quad . \tag{32}$$

It becomes clear also why and how various other forms of energy contribute to the total mass-energy-complexity: for example, a "binding", attractive interaction potential partially limits the chaotic motion liberty in the system and therefore makes a negative contribution to its mass-energy, while a repulsive interaction can only increase dynamic randomness and therefore contributes positively to the chaoticity-driven inertia.

## 5. Causal quantization of complex electron dynamics and wave equations

If we want to obtain an adequate description of the intrinsic wave-particle duality of the quantum beat process, then the above results should be completed by the explicitly wave-like, delocalised type of formal presentation accounting for the causally probabilistic wave field of the field-particle in its extended phase that corresponds to the physically real "intermediate" realisation through which the system performs its permanent transitions (or "quantum jumps") between the "regular" realisations of localised "virtual soliton" (section 3). That delocalised, partially ordered, dynamically maintained structure of the e/m protofield interacting with the gravitational protofield is described by the *physically based wavefunction*, $\Psi$, which arises from the initial "state-function" of the interacting protofields as its special, "intermediate" realisation (section 2) and is the causal extension of the canonical wavefunction introduced by Erwin Schrödinger (see ref. [24]). The essential difference between the two is due to the dynamically complex, chaotic physical nature of the extended version, which definitely resolves the basic difficulties of the canonical description around its *imposed* probabilistic "interpretation" of $\Psi$ by preserving the *direct* relation of the mathematical function $\Psi$ to the physically real, *dynamically probabilistic* wave field of the quantum beat process and thus causally justifying "Born's probability rule", which now explains $|\Psi|^2$ as the probability of emergence of the localised particle-realisation as a result of the dynamical auto-squeeze of the causal wavefunction (sections 2, 3) [14,15,24]. We resolve also the related problem of the abstract "configurational space" by providing the physical basis for the mathematical "coordinate", or "configuration", on which the canonical wavefunction depends: the causal wavefunction is defined on the *physically real* space of *dynamically emerging* centres of reduction, or "point"-realisations, the latter providing the causal version of system "configurations" (that can have, in principle, any "shape").

We see also that the interaction-driven *essential nonlinearity* of the particular system of real physical protofields, originating simply from their *unreduced interaction* and leading to the dynamic multiplicity of redundant state-realisations, permits us to replace a rather sophisticated and ambiguous combination of *several* components of an additional, specially introduced "quantum field" in the original version of the "double solution" [9] by the *single*, causally obtained wavefunction representing the *real* and *unifying* dynamical state of the participating *physical* fields. This state can be presented (section 4) as a naturally emerging dynamic combination of the regular, averaged wave-field structure (the de Broglie *matter* wave, now provided with the causally specified, really material basis) and the irregular wave field (with a characteristic structure of the size of the Compton wavelength, $\lambda_C = h/m_0 c$ [14]).

In order to obtain the delocalised, wavefunctional description in the described dual unity with the localised, corpuscular behaviour, we can suppose that the corresponding quantities, wavefunction $\Psi$ and action-complexity $\mathcal{A}$, are entangled in a single quantity of *wave action*, $\mathcal{A}_\Psi$, representing the *total* dynamic complexity of the quantum beat process, in all its aspects:

$$\mathcal{A}_\Psi = \mathcal{A} \Psi . \qquad (33)$$

The wave action is close to the action-complexity $\mathcal{A}$ of the virtual soliton motion, but explicitly takes into account the existence of the dynamically related extended system state (realisation). The physical basis of this relation between $\mathcal{A}_\Psi$, $\mathcal{A}$, and $\Psi$ becomes more transparent if we consider its change during one cycle of reduction-extension. The change of $\mathcal{A}_\Psi$ should be equal to zero, since the total complexity remains constant according to the complexity conservation law [14] (here it expresses the permanence of quantum beat in the isolated system of coupled protofields). Indeed, if the cycle starts from the extended state, then it should end up with the *same* state, since there is *only one* intermediate realisation, "bonding together" all the "regular", localised realisations of the virtual soliton. Therefore

$$\Delta \mathcal{A}_\Psi = \mathcal{A}\Delta\Psi + \Psi\Delta\mathcal{A} = 0 , \qquad (34a)$$

or

$$\Delta\mathcal{A} = -h\frac{\Delta\Psi}{\Psi} , \qquad (34b)$$

since the characteristic value of action $\mathcal{A}$ during each cycle is equal to a fixed portion of $h$ (for example, for the harmonic oscillation profile it is equal to $\hbar = h/2\pi$). The latter quantity is then additionally multiplied by the imaginary unit, i, which does not change the physical sense of the above duality expression and accounts for the difference between wave and corpuscular states in wave description by complex numbers:

$$\Delta\mathcal{A} = -\mathrm{i}\hbar\frac{\Delta\Psi}{\Psi} . \qquad (34c)$$

The causally substantiated version of the differential, "Dirac" quantization rules is then obtained by using the above definitions of momentum, eq. (21), and energy, eq. (24):

$$p = \frac{\Delta\mathcal{A}}{\Delta x} = -\frac{1}{\Psi}\mathrm{i}\hbar\frac{\partial\Psi}{\partial x} , \quad p^2 = -\frac{1}{\Psi}\hbar^2\frac{\partial^2\Psi}{\partial x^2} , \qquad (35)$$

$$E = -\frac{\Delta\mathcal{A}}{\Delta t} = \frac{1}{\Psi}\mathrm{i}\hbar\frac{\partial\Psi}{\partial t} , \quad E^2 = -\frac{1}{\Psi}\hbar^2\frac{\partial^2\Psi}{\partial t^2} , \qquad (36)$$

where the form of higher powers of $p$ and $E$ reflects the wave nature of $\Psi$ [14], and $x$ can be directly extended to the three-dimensional coordinate vector. We emphasize the unreduced complex-dynamical meaning of the causal quantization rules: in the basic form of eqs. (34) they express the quantum-beat cycle as a dynamically continuous entanglement of the virtual soliton "propagation", in the delocalised form of a "wave" during quantum jumps ($\Delta\Psi$), with its localised form appearance during reduction-extension ($\Delta\mathcal{A}$). It is therefore natural that quantization rules can be considered as another expression of the causal definition of space and time structure (cf. eqs. (21), (24)), emerging within our extended version of the double solution and entering now into the delocalised (undulatory) description of complex field-particle dynamics.

Inserting now the quantization expressions into a version of eq. (30),

$$E = m_0 c^2\sqrt{1-\frac{v^2}{c^2}} + \frac{p^2}{m} , \quad \text{or} \quad mE = m_0 c^2 + p^2 , \qquad (30'')$$

we get the dual, nonlocal formulation in the form of *wave equation* for $\Psi$ which can be considered as the simplest form of both Klein-Gordon and Dirac wave equations:

$$\mathrm{i}\hbar m\frac{\partial\Psi}{\partial t} + \hbar^2\frac{\partial^2\Psi}{\partial x^2} - m_0^2 c^2\Psi = 0 , \qquad (37a)$$

$$-\frac{\hbar^2}{c^2}\frac{\partial^2\Psi}{\partial t^2} + \hbar^2\frac{\partial^2\Psi}{\partial x^2} - m_0^2 c^2\Psi = 0 , \qquad (37b)$$

$$\frac{\partial^2\Psi}{\partial t^2} - c^2\frac{\partial^2\Psi}{\partial x^2} + \omega_0^2\Psi = 0 , \qquad (37c)$$

where $\omega_0 \equiv m_0 c^2/\hbar = 2\pi\nu_0$ is the rest-frame "circular" frequency of the quantum beat (which accounts actually for the spin vorticity twist [14]). More elaborated forms of wave equation taking into account particle interactions with external field and gravitational effects can be obtained with the help of the same causal quantization procedure [14], providing thus the realistic, complex-dynamical substantiation for the formally identical canonical versions. In the nonrelativistic limit they are reduced to the Schrödinger equation, but provided now with the physically consistent, causally complete origin and interpretation.

We can also obtain the Schrödinger equation by directly applying causal quantization rules, eqs. (35), (36), to the nonrelativistic limit of the energy-momentum relation of eq. (30´´) written for the field-particle in the external potential $V(x,t)$ (the latter causally accounts for the dynamically "folded", "potential" form of complexity, or "dynamical information" [14]):

$$E = \frac{p^2}{2m_0} + V(x,t) \;\;\rightarrow\;\; i\hbar\frac{\partial \Psi}{\partial t} = -\frac{\hbar^2}{2m_0}\frac{\partial^2 \Psi}{\partial x^2} + V(x,t)\Psi(x,t) \;. \tag{38}$$

It is not difficult to understand [14] that for binding potentials eq. (38) can be satisfied only for those discrete configurations of $\Psi(x,t)$ that correspond to integer numbers of the same action-complexity quantum, $h$, that describes one quantum-beat cycle, or "system realisation change", which explains the famous quantum-mechanical "energy-level discreteness" by the same, *universal* discreteness (*dynamic quantization*) of the underlying interaction process.

## 6. Physical origin of fundamental interactions, their unification, and complex-dynamical basis of general relativity effects

The unreduced description of reality within the emerging dynamic redundance paradigm and universal complexity concept (section 2) provides the naturally "cosmological", evolutionary picture of the world dynamics, where the existing entities are explicitly obtained in full agreement with their actual appearance in the universe formation processes (this is a manifestation of the explicit creativity of our analysis, as opposed to its invariable absence in the conventional, unitary theory, cf. section 1). In particular, it becomes clear that the unreduced interaction of two extended protofields leads to emergence of many local dynamically multivalued processes of quantum beat, each of them representing a well-specified physical structure of the elementary particle. In agreement with the basic property of dynamic multivaluedness, there can be many different elementary particles, each of them corresponding to a particular EP realisation (specified mainly by the potential well parameters). The appearing particle-processes modify local properties (average tension, density) of the protofields, which leads to interaction between particles through each of the protofields. The increasing protofield tension leads also to saturation of the total number of particles (and thus matter) in the universe, for the given protofield properties and magnitude of their attraction to each other. Due to the fundamentally self-consistent character of the unreduced EP dynamics (section 2), we obtain thus a *self-tuning universe*, which resolves a whole series of stagnating problems in the conventional theory.

Limiting ourselves here only to a brief outline of the obtained results (see refs. [14-16,21] for more details), we note as one of the most physically transparent results the explanation of the origin, total number (four), and basic properties of the fundamental interaction types between the elementary particles. Namely, the e/m and "weak" interaction forces are transmitted through the e/m protofield material, where the e/m interactions correspond to the "long-range" interaction type (analogous to "deformational" interaction between density perturbations in continuous media), while the weak interaction force originates from the direct, short-range interaction between the e/m protofield constituents. Similarly, the gravitational and "strong" interaction forces are transmitted through the gravitational protofield/medium (represented most probably by a dense quark condensate), where the universal gravitation (thus causally explained) appears as the long-range interaction mode, while the strong interaction results from the short-range, "contact" interaction between proto-quarks (gravitational protofield elements). It is extremely important that the four fundamental interactions thus obtained are intrinsically and permanently unified within the quantum beat dynamics, especially of the heaviest elementary particles. We also obtain the quantum (dynamically quantized) origin of all fundamental interactions, including gravity, which resolves another stagnating problem of the unitary theory (but the result is different from the conventional attempts and expectations of formal "quantization of gravity" and will be practically important only at extremely small space and time scales [14,16]). Various intrinsic properties of the fundamental particle interactions also naturally emerge in our description, such as the direct relation between inertial and gravitational manifestations of the quantum beat process (the extended "principle of equivalence"), and the origin, universality, and two "opposite" species of the quantized electric charge.

Note also the essential difference between the *postulated* fixed, "geometrical" (abstract) description of gravity in the conventional "general relativity" and our causally derived, purely *dynamical*, and *naturally quantized* gravitation that produces, however, the same "relativistic" modification of space and time, but now causally related to their clearly specified physical origin (where space and time emerge as *qualitatively different*, only *dynamically* related entities) [14,16]. In the unreduced description of gravity, the nonuniform tension of the gravitational medium is associated with the corresponding inhomogeneous distribution of individual quantum beat processes and can be expressed by the respective "gravitational potential", rather than a formal "metric" of the abstract space-time geometry. Thus, for the case of static gravitational field (cf. ref. [23]) one obtains, instead of eq. (19):

$$h\nu_0(x) = m_0 c^2 \sqrt{g_{00}(x)} \ , \tag{39}$$

where the classical metric, $g_{00}(x)$, should be understood rather as expression containing the gravitational potential that characterises inhomogeneous distribution of tension in the gravitational medium (for the case of weak fields, $g_{00}(x) = 1 + 2\phi_g(x)/c^2$, where $\phi_g(x)$ is the classical gravitational field potential [23]). Since $\nu_0(x)$ determines the causal "time flow" (section 4) and $\phi_g(x)$ has the negative sign ($g_{00}(x) < 1$) corresponding to the universal gravitational attraction, the above equation substantiates the causal version of the "relativistic time retardation" in the gravitational field. This natural involvement of gravity with the dynamically quantized behaviour of micro-objects can be further specified [14,16], but we shall not present here all the details, being interested rather in the fundamental unification of quantum and relativistic behaviour, originating from the unreduced dynamic complexity of the elementary field-particles. The most fundamental, and always present, evidence of this *unique causal origin* of the quantum "mysteries" and relativistic "paradoxes", the unreduced dynamic complexity (multivaluedness) of the underlying interaction process, is provided by the intrinsically unified mass-energy-complexity, determined by the rate (intensity) of the chaotic quantum beat process and giving quantum behaviour, inertia, gravity, and relativistic dynamics (see eqs. (18)-(19), (23)-(26), (30)-(32), (36)-(38), (39)).

# 7. Classical behaviour emergence and the unified hierarchy of complex dynamics of the world

The extended causality should include dynamic unification with higher-level systems and behaviour, in the form of their *natural emergence* from the described level of elementary field-particles. The next higher sublevel of complexity results indeed from the essentially quantum dynamics, in the form of dynamically redundant and therefore chaotic (complex) behaviour of the interacting field-particles. Dynamic complexity here can take the form of either *true* quantum (Hamiltonian) chaos (in the absence of dissipation), or causally extended quantum measurement (slightly dissipative, open-system dynamics) [14,17]. The latter can be considered as a complex-dynamical particle interaction process (cf. section 2), but involving, due to dissipativity, formation of a transient quasi-bound state between two (or several) elementary field-particles whose virtual solitons perform, during a transiently short period, their chaotic "quantum jumps" in a close vicinity of each other, thus temporary reducing the long-range "wave properties" of the "measured" particle.

When the (attractive) interaction magnitude is large enough, such transient bound state of the field-particles can become a stable (permanent) bound state (like that of an atom), and a *new level of complexity* emerges, that of the *classical behaviour*. Indeed, due to the sufficiently strong attraction within the bound state of elementary field-particles their virtual solitons should always remain close to each other while performing their spatially *random* jumps, which limits the probability of their larger "correlated walks" to low, exponentially decreasing values and results in a well-localised, confined, or trajectorial type of chaotic dynamics. This (complex-) dynamically emerging classical character of the elementary (and larger) bound systems is quite different from both canonical *semi-classical* behaviour that need not be localised, and canonical (ill-defined) notion of classicality, since our elementary classical system need not be "macroscopic" or "dissipative" (open to contacts with an "environment" that produces "decoherence"), but is always *internally*, dynamically chaotic (even though external interactions can *quantitatively* modify localisation and the ensuing classicality).

The hierarchical complexity development continues in the same general fashion to all the higher levels: the quasi-localised trajectory of a classical, bound system becomes dynamically irregular, and thus again delocalised, due to any real interaction(s) with other classical system(s), then bound, "condensed" (many-particle) systems of higher levels are formed, etc. It is important that the same qualitative concept of dynamic redundance and the related quantitative description remain valid at any level of (irreducibly complex) reality and are represented by the localised, trajectorial formalism (Hamilton-Lagrange equation generalising the ordinary Hamilton-Jacobi equation) and delocalised, state-functional formalism (generalised Schrödinger equation) [14,24], the latter extending various canonical equations for quantum and classical "density matrix" and "distribution function". We see now that the consistent, causal explanation of quantum phenomena is attained *together with* the *truly* consistent, and *essentially* different from canonical, causally *extended* description of the classical, macroscopic behaviour, within the same concept of the unreduced dynamic complexity. All the "specific" strangeness of quantum behaviour is due simply to the fact that it is observed at several *lowest*, naturally coarse-grained levels of complexity, where dissipation (interaction with other levels) is strongly suppressed and the universal process of chaotic realisation change appears in its unveiled, "explicitly complex-dynamical" (dynamically multivalued) form which only *seems* to be weird with respect to unrealistically simplified, dynamically single-valued approach of the canonical, unitary theory.